\documentclass[11pt,letter]{article}
\usepackage{amssymb, fullpage}
\newtheorem{theorem}{Theorem}

\newtheorem{definition}[theorem]{Definition}

\newtheorem{remk}[theorem]{Remark}
\newtheorem{exmp}[theorem]{Example}

\def\FullBox{\hbox{\vrule width 8pt height 8pt depth 0pt}}

\def\qed{\ifmmode\qquad\FullBox\else{\unskip\nobreak\hfil
\penalty50\hskip1em\null\nobreak\hfil\FullBox
\parfillskip=0pt\finalhyphendemerits=0\endgraf}\fi}

\def\qedsketch{\ifmmode\Box\else{\unskip\nobreak\hfil
\penalty50\hskip1em\null\nobreak\hfil$\Box$
\parfillskip=0pt\finalhyphendemerits=0\endgraf}\fi}

\newcommand{\poly}{{\mathrm{poly}}}

\begin{document}
\title{New Algorithms for Heavy Hitters in Data Streams
\footnote{A preliminary version of this paper appeared 
as an invited paper in ICDT 2016.}}
\author{David P. Woodruff\\ IBM Research Almaden\\dpwoodru@us.ibm.com}
\date{}
\maketitle
\begin{abstract}
An old and fundamental problem in databases and data streams is that of finding the heavy hitters, also known as the top-$k$, most popular items, frequent items, elephants, or iceberg queries. There are several variants of this problem, which quantify what it means for an item to be frequent, including what are known as the $\ell_1$-heavy hitters and $\ell_2$-heavy hitters. There are a number of algorithmic solutions for these problems, starting with the work of Misra and Gries, as well as the CountMin and CountSketch data structures, among others. 

In this survey paper, accompanying an ICDT invited talk, we cover several recent results developed in this area, which improve upon the classical solutions to these problems. In particular, with coauthors we develop new algorithms for finding $\ell_1$-heavy hitters and $\ell_2$-heavy hitters, with significantly less memory required than what was known, and which are optimal in a number of parameter regimes. 
\end{abstract}

\section{The Heavy Hitters Problem}
A well-studied problem in databases and data streams is that of finding the heavy hitters, also known as the top-$k$, most popular items, frequent items, elephants, or iceberg quries. These can be used for flow identification at IP routers \cite{ev03}, in association rules and frequent itemsets \cite{as94,hpy00,hid99,son95,toi96}, 
and for iceberg queries and iceberg datacubes \cite{br99,fsgmu98,HPDW01}. We refer the reader to the survey \cite{cormode2008finding}, which presents an overview of known algorithms for this problem, from both theoretical and practical standpoints.

There are various different flavors of guarantees for the heavy hitters problem. We start with what is known as the $\ell_1$-guarantee:

\begin{definition}($\ell_1$-{\bf $(\epsilon, \phi)$-Heavy Hitters Problem})
In the $(\epsilon, \phi)$-Heavy Hitters Problem, we are given parameters
$0 < \epsilon < \phi < 1$, as well as  
a stream $a_1, \ldots, a_m$ of items $a_j \in \{1, 2, \ldots, n\}.$ 
Let $f_i$ denote the number of occurrences of item $i$, i.e., its 
frequency.  
The algorithm should make one pass over the stream and at the
end of the stream output a set $S \subseteq \{1, 2, \ldots, n\}$
for which if $f_i \geq \phi m$, then $i \in S$, while if 
$f_i \leq (\phi - \epsilon)m$, then $i \notin S$. Further, for
each item $i \in S$, the algorithm should output an estimate
$\tilde{f}_i$ of the frequency $f_i$ which satisfies
$|f_i - \tilde{f}_i| \leq \epsilon m$. 
\end{definition}

We are
interested in algorithms which use as little space (i.e., memory) in bits
as possible to solve the $\ell_1$-{\bf $(\epsilon, \phi)$-Heavy Hitters Problem}.
We allow the algorithm to be randomized and
to succeed with probability at least $1-\delta$, for $0<\delta<1$.
We do not make any assumption on the ordering of the stream
$a_1, \ldots, a_m$. This
is desirable, as often in applications one cannot assume a best-case or even
a random order. We will assume $m$ is known in advance, though many of the algorithms below (including ours)
can deal with unknown $m$. 
%We note that while the problem still makes sense for any $\phi > \epsilon$,
%it is well-known that an $\Omega(n)$ space lower bound exists when $\phi$ is very close to $\epsilon$,
%e.g., if $\phi = \epsilon + 1/n$. Indeed, this follows via a reduction from communication complexity,
%which is a standard method for proving lower bounds in data streams. In particular, a reduction from
%the so-called {\sf INDEX} problem is readily apparent - we refer the reader to \cite{kremer1999randomized} 
%for more details of the communication problem (see, e.g., \cite{r15} for a recent survey discussing the {\sf INDEX} problem).  

The first algorithm for the $\ell_1$-{\bf $(\epsilon, \phi)$-Heavy Hitters Problem}
was given by Misra and Gries \cite{misra82}, who achieved $O(\epsilon^{-1} \log n)$
bits of space for any $\phi > 2\epsilon$. 
This algorithm was rediscovered
by Demaine et al. \cite{demaine2002frequency}, and again by Karp et al. \cite{karp2003simple}. 
Other than these 
algorithms, which are deterministic, there are a number of randomized
algorithms, such as the CountSketch \cite{charikar2004finding}, Count-Min sketch \cite{cormode2005improved}, sticky sampling \cite{mm02},
lossy counting \cite{mm02}, space-saving \cite{MetwallyAA05}, sample and hold \cite{ev03}, multi-stage bloom filters
\cite{cfm09}, and sketch-guided sampling \cite{kx06}. Berinde
et al. \cite{bics10} show that using $O(k \epsilon^{-1} \log(mn))$ bits of space, 
one can achieve the stronger guarantee
of reporting, for each item $i \in S$, $\tilde{f}_i$ with
$|\tilde{f}_i - f_i| \leq \frac{\epsilon}{k} F^{res(k)}_1$, 
where $F^{res(k)}_1 < m$ denotes
the sum of frequencies of items in $\{1, 2, \ldots, n\}$ excluding the frequencies
of the $k$ most frequent items. This is particularly useful when there are only a few
large frequencies, since then the error $\frac{\epsilon}{k} F^{res(k)}_1$ will depend
only on the remaining small frequencies. 

While the $\ell_1$-heavy hitters have a number of applications, there is also a sometimes
stronger notion known as the $\ell_2$-heavy hitters, which we now define. 

\begin{definition}($\ell_2$-{\bf $(\epsilon, \phi)$-Heavy Hitters Problem})
In the $(\epsilon, \phi)$-Heavy Hitters Problem, we are given parameters
$0 < \epsilon < \phi < 1$, as well as  
a stream $a_1, \ldots, a_m$ of items $a_j \in \{1, 2, \ldots, n\}.$
Let $f_i$ denote the number of occurrences of item $i$, i.e., its 
frequency.  Let $F_2 = \sum_{i=1}^n f_i^2$. 
The algorithm should make one pass over the stream and at the
end of the stream output a set $S \subseteq \{1, 2, \ldots, n\}$
for which if $f_i^2 \geq \phi F_2$, then $i \in S$, while if 
$f_i^2 \leq (\phi - \epsilon)F_2$, then $i \notin S$. Further, for
each item $i \in S$, the algorithm should output an estimate
$\tilde{f}_i$ of the frequency $f_i$ which satisfies
$|f_i - \tilde{f}_i| \leq \epsilon \sqrt{F_2}$. 
\end{definition}

One of the algorithms for $\ell_1$-heavy hitters mentioned above, the CountSketch \cite{ccf04}, refined in \cite{tz12}, 
actually solves the $\ell_2$-{\bf $(\epsilon, \phi)$-Heavy Hitters Problem}. Notice
that this guarantee can be significantly stronger than the aforementioned $\ell_1$-guarantee that $f_i \geq \epsilon m$.  
Indeed, if $f_i \geq \phi m$, then $f_i^2 \geq \phi^2 m^2 \geq \phi^2 F_2$. So, an algorithm
for finding the $\ell_2$-heavy hitters, with $\phi$ replaced by $\phi^2$, will find all items satisfying the $\ell_1$-guarantee with 
parameter $\phi$. On the other hand,
given a stream of $n$ distinct items in which $f_{i^*} = \sqrt{n}$ for an $i^* \in [n] = \{1, 2, 3, \ldots, n\}$, 
yet $f_i = 1$ for all $i \neq i^*$,  
an algorithm satisfying the $\ell_2$-heavy hitters guarantee will identify item $i$ with constant $\phi$, but
an algorithm which only has the $\ell_1$-guarantee would need to set $\phi = 1/\sqrt{n}$, therefore using
$\Omega(\sqrt{n})$ bits of space. In fact,
$\ell_2$-heavy hitters are in some sense the best one can hope for with a small amount of space in a data stream,
as it is known for $p > 2$ that finding those $i$ for which $f_i^p \geq \phi F_p$ requires $n^{1-2/p}$ bits of space even for 
constant $\phi$ \cite{bjks04,cks03}. 

The $\ell_2$-heavy hitter algorithms of \cite{ccf04,tz12} have broad applications in compressed sensing \cite{glps10,mp14,p11} and 
numerical linear algebra \cite{bn13,cw13,mm13,nn13}, and are often used as a subroutine in other data stream
algorithms, such as  $\ell_p$-sampling \cite{ako11,jst11,mw10}, cascaded aggregates \cite{jw09}, and 
frequency moments \cite{bgks06,iw05}. 

Given the many applications of heavy hitters, it is natural to ask
what the best space complexity for them is. For simplicity of presentation, we make the common assumption that the
stream length $m$ is polynomially related to the universe size $n$. 

It is clear that for constant $\epsilon$ and $\phi$, that there is 
an $\Omega(\log n)$ bit lower bound, as this is just the number of bits needed to specify the identity of the heavy hitter.

For constant $\epsilon$, given the aforementioned results, this is actually tight for the
$\ell_1$-{\bf $(\epsilon, \phi)$-Heavy Hitters Problem}. The main focus then, for the 
$\ell_1$-{\bf $(\epsilon, \phi)$-Heavy Hitters Problem} is on obtaining tight bounds as a function of $\epsilon$ and $\phi$. 

On the other hand, for the $\ell_2$-{\bf $(\epsilon, \phi)$-Heavy Hitters Problem}, even for constant $\epsilon$ and $\phi$,
the best previous algorithms of \cite{ccf04} and the followup
\cite{tz12} achieve $\Theta(\log^2 n)$ bits of space. It is known that if one allows deletions
in the stream, in addition to insertions, then $\Theta(\log^2 n)$ bits of space is optimal
\cite{bipw11,jst11}. However, in many cases we just have a stream of insertions, such as in the model
studied in the seminal paper of Alon, Matias, and Szegedy \cite{ams99}. Thus, for the 
 $\ell_2$-{\bf $(\epsilon, \phi)$-Heavy Hitters Problem}, our focus will be on the regime of constant $\epsilon$ and $\phi$ and on understanding the dependence on $n$.

There are a number of other desirable properties one would want out of a heavy hitters algorithm. 
For instance, one is often also interested in minimizing the {\it update time}
and {\it reporting time} of such algorithms. Here, the update time
is defined to be the time the algorithm needs to update its data
structure when processing a stream insertion. The reporting time is the
time the algorithm needs to report the answer after having 
processed the stream. In this article we will focus primarily on the space complexity. For other very interesting recent work on improving the reporting time in a stream of insertions and deletions, see \cite{lnnt16}. The results in this
survey are focused on a stream of insertions only (for which, as mentioned above, smaller space bounds are possible).   

\section{Our Recent Results}
In several recent works \cite{bdw16,bcinww16,bciw15}, we significantly improve known algorithms for finding both $\ell_1$-heavy hitters as well as $\ell_2$-heavy hitters. 
For many settings of parameters, our algorithms are optimal. 

\subsection{$\ell_1$-Heavy Hitters}
In joint work with Bhattacharyya and Dey \cite{bdw16}, we improve upon the basic algorithm of Misra and Gries 
\cite{misra82} 
for the $\ell_1$-{\bf $(\epsilon, \phi)$-Heavy Hitters Problem}, the latter achieving $O(\epsilon^{-1} \log n)$ bits of space for any 
$\phi \geq 2\epsilon$. There are two algorithms of \cite{bdw16}, the first
a bit simpler and already achieving a large improvement over 
\cite{misra82}, and the second an optimal algorithm. We first discuss
the first algorithm.

We first recall the algorithm of Misra and Gries. That algorithm initializes a table of $1/\epsilon + 1$ pairs of $(v, c)$ to $(\bot, 0)$, where
$v$ is an element in the universe $\{1, 2, \ldots, n\} \cup \bot$, and $c$ is a non-negative integer. When receiving a new 
stream insertion $a_i$, the algorithm checks if $v = a_i$ for some $(v,c)$ pair in the table. If so, it replaces $(v,c)$ with
$(v, c+1)$. Otherwise, if there is a $(v,c)$ in the table with $v = \bot$, then the algorithm replaces that $(v,c)$ pair
with $(a_i, 1)$. If neither of the previous two cases hold, the algorithm takes each $(v,c)$ pair in the table, and replaces it with
$(v,c-1)$. If $c-1 = 0$, then the corresponding $v$ is replaced with $\bot$. 

Note that the algorithm, as described in the previous paragraph, naturally can be implemented using $O(\epsilon^{-1} \log n)$ bits of space
(recall we assume the stream length $m$ and the universe size $n$ are polynomially related, so $\log m = \Theta(\log n)$). Moreover, a
nice property is that the algorithm is deterministic. 

For the correctness, note that if an item $i$ occurs $f_i \geq 2\epsilon m$ times, then it will appear
in the table at the end of the stream. Indeed, notice that for each occurrence of $i$ in the stream, if it is not included in the table
via the operation of replacing a pair $(i,c)$ with $(i,c+1)$ for some value of $c$, or replacing a pair $(\bot, 0)$ with $(i, 1)$, then this means
that there were at least $1/\epsilon+1$ stream updates that were removed from the table upon seeing this occurrence of $i$, since each
counter $c$ for each $(v,c)$ pair in the table is decremented by $1$. We can therefore
charge those stream updates to this occurrence of $i$. Moreover,
if $(i,c)$ is in the table for some value of $c$ and is replaced with $(i, c-1)$ or $(\bot, 0)$, this means we can charge 
at least $1/\epsilon$ stream updates to items not equal to $i$ to this occurrence of $i$. Since we are charging distinct stream updates for each occurrence of $i$, we have the relationship that $f_i \cdot (1/\epsilon) \leq m$, which is a contradiction to $f_i \geq 2\epsilon m$. 
Therefore, $i$ will occur in a pair in the table at the end of the stream. The same analysis in fact implies that at most $\epsilon m$ occurrences
of $i$ will not be accounted for in the table at the end of the stream, which means that for the $(i,c)$ pair in the table, we have
$f_i \geq c \geq f_i - \epsilon m$. This latter guarantee enables us to solve the 
$\ell_1$-{\bf $(\epsilon, \phi)$-Heavy Hitters Problem} for any $\phi \geq 2 \epsilon$. 

One shortcoming of the algorithm above is that if $\phi$ is much larger than $\epsilon$, say $\phi$ is constant, then the above algorithm still
requires $O(\epsilon^{-1} \log n)$ bits of space, that is, it is insensitive to the value of $\phi$. Consider for instance, the case when
$\epsilon = 1/\log n$ and $\phi = 1/10$, so one wants a very high accuracy estimate to each of the item frequencies for items occurring at least 10\% of the time. The above
algorithm would use $O(\log^2 n)$ bits of space for this problem. In this case, the only known lower bound is $\Omega(\log n)$ bits, which just
follows from the need to return the identities of the heavy hitters. Is it possible to improve this $O(\log^2 n)$ bits of space upper bound?

This is precisely what we show in \cite{bdw16}. Here we sketch how to achieve a bound of $O((1/\phi) \log n + (1/\epsilon) \log (1/\epsilon))$ bits of space 
and refer to \cite{bdw16} for further optimizations as well as extensions to related problems. Note that this translates to a space bound
of $O(\log n \log \log n)$ bits for the above setting of parameters. 

The first observation is that if we randomly sample $r = \Theta(1/\epsilon^2)$ stream updates, then with probability 99\%, 
simultaneously 
for every universe item $i$, if we let
$\hat{f}_i$ denote its frequency among the samples, and $f_i$ its frequency in the original stream, then we have 
$$\left |\frac{\hat{f}_i}{r} - \frac{f_i}{m} \right | \leq \frac{\epsilon}{2}.$$
This follows by Chebyshev's inequality and a union bound. Indeed, consider a given $i \in [n]$ with frequency $f_i$ and suppose
we sample each of its occurrences pairwise-independently with probability $r/m$, for a parameter $r$. Recall that pairwise
independence here implies that any single occurrence is sampled with probability $r/m$ and any 
two occurrences are jointly sampled with probability exactly $r^2/m^2$, though we do not impose any constraints on the joint distribution
of any three or more samples. Also, a pairwise independent hash function can be represented with only $O(\log n)$ bits of space. 
Then the expected number ${\bf E}[\hat{f_i}]$ of sampled occurrences is 
$f_i \cdot r/m$ and the variance ${\bf Var}[\hat{f_i}]$ is $f_i \cdot r/m (1-r/m) \leq f_ir/m$ (here we use pairwise independence to conclude the same
variance bound as if the samples were fully independent). Applying Chebyshev's inequality, 
$$
\Pr \left [\left | \hat{f_i} - {\bf E}[\hat{f_i}] \right | \geq \frac{r \epsilon}{2} \right ] \leq \frac{{\bf Var}[\hat{f_i}]}{(r \epsilon/2)^2}
\leq \frac{4f_ir}{m r^2 \epsilon^2}.$$
Setting $r = \frac{C}{\epsilon^2}$ for a constant $C > 0$ makes this probability at most $\frac{4f_i}{C m}$. By the union bound, if we
sample each element in the stream independently with probability $\frac{r}{m}$, then the probability there exists an $i$ for which
$|\hat{f_i} - {\bf E}[\hat{f_i}]| \geq \frac{r \epsilon}{2}$ is at most $\sum_{i = 1}^n \frac{4 f_i}{C m} \leq \frac{4}{C}$, which for
$C \geq 400$ is at most $\frac{1}{100}$, as desired. 

After sampling so that the stream length is reduced to $O(1/\epsilon^2)$, it follows that the number of distinct items in the stream is
also $O(1/\epsilon^2)$, and therefore if we hash the item identifiers to a universe of size $O(1/\epsilon^4)$, by standard arguments with
probability 99\% the items will be perfectly hashed, that is, there will be no collisions. This follows even with a pairwise-independent
hash function $h$. The high level idea then is to run the algorithm of Misra and Gries, but the pairs $(v,c)$ correspond to the {\it hashed}
item identity and the {\it count in the sampled stream}, respectively. Notice that it takes only $O(\log(1/\epsilon))$ bits to represent
such pairs and so the algorithm of Misra and Gries would take $O(\epsilon^{-1} \log(1/\epsilon))$ bits of space. 

However, we still want
to return the actual item identifiers! To do this, we maintain a parallel data structure containing actual item identifiers
in $[n]$, but the data structure only contains $O(1/\phi)$ items. In particular, these item identities correspond to the items $v$
for which $(h(v), c)$ is stored in the algorithm of Misra and Gries, for which the $c$ values are largest. Namely, the items with
top $1/\phi$ $c$-values have their actual identities stored. This can be maintained under stream insertions since given a new stream
update, one has the actual identity in hand, and therefore can appropriately update the identities of the items with top $O(1/\phi)$ counts. 
Moreover, when we subtract one from all counters in the algorithm of Misra and Gries, the only thing that changes in the top $O(1/\phi)$
identities is that some of them may now have zero frequency, and so can be thrown out. Thus, we can always maintain the actual top
$O(1/\phi)$ identities in the original (before hashing) universe.

The second algorithm of \cite{bdw16} achieves an optimal
$O(\epsilon^{-1} \log \phi^{-1} + \phi^{-1} \log n + \log \log m)$ bits of 
space. The algorithm can be seen as an extension of our first algorithm.
The idea of the optimal 
algorithm, as in our first algorithm, is to have a list of the top
$O(1/\phi)$-heavy hitters with exact identities, and to use a separate data structure to approximate their individual
frequencies up to $\epsilon \cdot m$. In the earlier algorithm, this was an accompanying Misra-Gries data
structure on the hashed universe identities and sample count values; 
in the new one we optimize this data structure
to use $O(\epsilon^{-1} \log \phi^{-1})$ bits instead of the earlier $O(\epsilon^{-1} \log \epsilon^{-1})$ bits. 
We have $O(1/\epsilon)$ counts, as before, but now in each count we spend $O(1)$ bits on average. We also eliminate
the need to maintain hashed identities in the earlier algorithm by partitioning the items into $O(1/\epsilon)$
buckets using a hash function and maintaining the approximate sum in each bucket. We note
that the counts need to be randomized, but in a different sense than probabilistic counters since we want them to 
achieve additive error $O(1/\epsilon)$ rather than the relative error guarantee of probabilistic
counters. We call these {\it accelerated
counters} since their relative error improves as the count gets larger. We are able
to compress the counts since they sum up to $O(1/\epsilon^2)$, which is the length of the sampled stream. Each count
is individually only correct with constant probability, so we have $O(\log(1/\phi))$ repetitions and take a median
across the repetitions to get a correct count for each of the $O(1/\phi)$ heavy hitters in our list. 

We refer the reader to \cite{bdw16} for further details about both algorithms. 

\subsection{$\ell_2$-Heavy Hitters}
In joint work with Braverman, Chestnut, and Ivkin \cite{bciw15}, we improve upon the CountSketch data structure 
\cite{ccf04} for the $\ell_2$-{\bf $(\epsilon, \phi)$-Heavy Hitters Problem}. To illustrate the algorithm of \cite{bciw15}, we consider $\epsilon$
and $\phi$ to be constants in what follows, and further, we suppose there is only a single $i^* \in [n]$ for which $f_{i^*}^2 \geq \phi F_2$
and there is no $i$ for which $(\phi-\epsilon) F_2 \leq f_i^2 < \phi F_2$. It is not hard to reduce to this case by first hashing into
$O(1)$ buckets (recall $\phi, \epsilon$ are constants for this discussion), since the $O(1/\phi)$ heavy hitters will go to separate
buckets with large constant probability (if, say, we have $\Omega(1/\phi^2)$ buckets). Thus, we focus on this case. In this case the CountSketch algorithm would use $\Theta(\log^2 n)$
bits of space, whereas in \cite{bciw15} we achieve $O(\log n \log \log n)$ bits of space, nearly matching the trivial $\Omega(\log n)$
bit lower bound. 

We first explain the CountSketch data structure. The idea is to assign each item $i \in [n]$ a random sign $\sigma(i) \in \{-1,1\}$. We
also randomly partition $[n]$ into $B$ buckets via a hash function $h$ 
and maintain a counter $c_j = \sum_{i \mid h(i) = j} \sigma(i) \cdot f_i$ in the $j$-th bucket. Then, to estimate any given frequency
$f_i$, we estimate it as $\sigma(i) \cdot c_{h(i)}$. Note that ${\bf E}[\sigma(i) \cdot c_{h(i)}] = {\bf E}[\sigma(i)^2 f_i + \sum_{j \neq i, h(j) = h(i)} f_j \sigma(j) \sigma(i)] = f_i$, using that ${\bf E}[\sigma(i) \sigma(j)] = 0$ for $i \neq j$. Moreover, by computing the
variance and applying Chebyshev's inequality, one has that $$|\sigma(i) \cdot c_{h(i)} - f_i| = O(\sqrt{F_2/B})$$ with probability
at least $9/10$. The intuitive explanation is that due to the random sign combination of remaining items in the same hash bucket as $i$,
the absolute value of this linear combination concentrates to the Euclidean norm of the frequency vector of these items. 
The idea then is to repeat this independently $O(\log n)$ times in parallel. Then we estimate $f_i$ by taking the
median of the estimates across each of the $O(\log n)$ repetitions. By Chernoff bounds, we have that with probability $1-1/n^2$, say,
the resulting estimate is within an additive $O(\sqrt{F_2/B})$ of the true frequency $f_i$. This then holds for every $i \in [n]$
simultaneously by a union bound, at which point one can then find the $\ell_2$-heavy hitters, if say, one sets $B = \Theta(1/\epsilon^2)$.

Notice that it is easy to maintain the CountSketch data structure in a data stream since we just need to hash the new item $i$ to the appropriate
bucket and add $\sigma(i)$ to the counter in that bucket, once for each of the $O(\log n)$ repetitions. 
The total space complexity of the CountSketch algorithm is $O(B \cdot \log^2 n)$, where the ``$B$'' is the number of hash buckets,
one $\log n$ factor is to store the counter in each bucket, and the other $\log n$ factor is for the number of repetitions. For constant
$\epsilon$ and $B = \Theta(1/\epsilon^2)$ this gives $O(\log^2 n)$ bits of space. It is also not hard to see that the CountSketch data structure can be maintained
in a stream with deletions as well as insertions, since given a deletion to item $i$, this just corresponds to subtracting $\sigma(i)$
from the bucket $i$ hashes to in each repetition. Moreover, as mentioned earlier, this $O(\log^2 n)$ space bound is optimal for streams with deletions. 

To give some intuition for our new algorithm, let $i^* \in [n]$ be the identity of the single $\ell_2$-heavy hitter that we wish to find.
Suppose first that $f_{i^*} \geq \sqrt{n} \log n$ and that $f_i \in \{0,1\}$ for all $i \in [n] \setminus \{i^*\}$. For the moment, we are
also going to ignore the issue of storing random bits, so assume we can store $\poly(n)$ random bits for free (which can be indexed into
using $O(\log n)$ bits of space). We will later sketch how to remove this assumption. As in the CountSketch algorithm, we again assign
a random sign $\sigma(i)$ to each item $i \in [n]$. Suppose we randomly partition $[n]$ into two buckets using a hash function 
$h:[n] \rightarrow \{1,2\}$, and correspondingly maintain two counters $c_1 = \sum_{i \mid h(i) = 1} \sigma(i) \cdot f_i$ and
$c_2 = \sum_{i \mid h(i) = 2} \sigma(i) \cdot f_i$. Suppose for discussion that $h(i^*) = 1$. A natural question is what the values $c_1$
and $c_2$ look like as we see more updates in the stream. 

Consider the values $c_1 - \sigma(i^*) \cdot f_{i^*}$ and $c_2$. Then, since all frequencies other than $i^*$ are assumed to be $0$ or $1$,
and since the signs $\sigma(j)$ are independent, these two quantities evolve as random walks starting at $0$ and incrementing by $+1$
with probability $1/2$, and by $-1$ with probability $1/2$, at each step of the walk. By standard theory of random walks (e.g., Levy's theorem), there is a constant $C > 0$ so that with probability at least $9/10$, 
simultaneously at all times during the stream we have that $|c_1 - \sigma(i^*) \cdot f_{i^*}|$ and $|c_2|$
are upper bounded by $C \sqrt{n}$. The constant of $9/10$, like typical constants in this paper, is somewhat arbitrary. This suggests
the following approach to learning $i^*$: at some point in the stream we will have that $f_{i^*} > 2 C \sqrt{n}$, and at that point
$|c_1| > C \sqrt{n}$, but then we know that $i^*$ occurs in the first bucket. This is assuming that the above event holds for the random
walks. Since we split $[n]$ randomly into two pieces, this gives us $1$ bit of information about the identity of $i^*$. If we were
to repeat this $O(\log n)$ times in parallel, we would get exactly the CountSketch data structure, which would use $\Theta(\log^2 n)$
bits of space. Instead, we get much better space by repeating $\Theta(\log n)$ times sequentially!

To repeat this sequentially, we simply wait until either $|c_1|$ or $|c_2|$ exceeds $C n^{1/2}$, at which point we learn
one bit of information about $i^*$. Then, we reset the two counters to $0$ and perform the procedure again. Assuming 
$f_{i^*} = \Omega(\sqrt{n} \log n)$, we will have $\Omega(\log n)$ repetitions of this procedure, each one succeeding independently
with probability $9/10$. By Chernoff bounds, there will only be a single index $i \in [n]$ which match a $2/3$ fraction of these
repetitions, and necessarily $i = i^*$. 

\subsubsection{Gaussian Processes}
In general we do not have $f_{i^*} = \Omega(\sqrt{n} \log n)$, nor do we have that $f_i \in \{0,1\}$
for all $i \in [n] \setminus \{i^*\}$. We fix both problems using the theory of Gaussian processes. 

\begin{definition}\label{def:gp}
A {\bf Gaussian process} is a collection $\{X_t\}_{t \in T}$ of random variables, for an index set $T$, for which every
finite linear combination of the random variables is Gaussian. 
\end{definition}
We assume ${\bf E}[X_t] = 0$ for all $t$, as this will suffice for our application. It then follows that the Gaussian
process is entirely determined by its covariances ${\bf E}[X_s X_t]$. This fact is related to the fact that a Gaussian
distribution is determined by its mean and covariance. The distance
function $d(s,t) = ({\bf E}[(X_s-X_t)^2])^{1/2}$ is then a pseudo-metric on $T$ (the only property it lacks of a metric
is that $d(s,t)$ may equal $0$ if $s \neq t$). 

The connection to data streams is the following. Suppose we replace the signs $\sigma(i)$ with standard normal
random variables $g(i)$ in our counters above, and consider a counter $c$ at time $t$, denoted $c(t)$, of the form  
$\sum_i g(i) \cdot f_i(t)$. Here $f_i(t)$ is the frequency of item $i$ after processing $t$ stream insertions. 
The main point is that $c(t)$ is a Gaussian process! Indeed, any linear combination of the $c(t)$ values for
different $t$ is again Gaussian since the sum of normal random variables is again a normal random variable. 

The reason we wish to make such a connection to Gaussian processes is the following powerful inequality called the
``chaining inequality''. 
\begin{theorem}(Talagrand \cite{t96}) \label{thm:chain}
Let $\{X_t\}_{t \in T}$ be a Gaussian process and let $T_0 \subseteq T_1 \subseteq T_2 \subseteq \cdots \subseteq T$ be
such that $|T_0| = 1$ and $|T_i| \leq 2^{2^i}$ for $i \geq 1$. Then,
$${\bf E} \left [\sup_{t \in T} X_t \right ] \leq O(1) \cdot \sup_{t \in T} \sum_{i \geq 0} 2^{i/2} d(t, T_i),$$
where $d(t, T_i) = \min_{s \in T_i} d(t, s)$. 
\end{theorem}
We wish to apply Theorem \ref{thm:chain} to the problem of finding $\ell_2$-heavy hitters. Let $F_2(t)$ be the value
of the second moment $F_2$ after seeing $t$ stream insertions. We now describe how to choose the sets $T_i$ in order
to apply the chaining inequality; the intuition is that we recursively partition the stream based on its $F_2$ value.

Let $a_t$ be the first stream update for which $F_2(m)/2 \leq F_2(t)$. Then $T_0 = \{t\}$. 
We then let $T_i$ be the set of $2^{2^i}$ times $t_1, t_2, \ldots, t_{2^{2^i}}$ in the stream for which $t_j$ is the
first point in the stream for which $j \cdot F_2(m)/2^{2^i} \leq F_2(t_j)$. Then, we have created a nested
sequence of subsets $T_0 \subseteq T_1 \subseteq T_2 \subseteq \cdots \subseteq T$ with $|T_0| = 1$ and
$|T_i| \leq 2^{2^i}$ for $i \geq 1$. 

We are now in position to apply Theorem \ref{thm:chain}. A straightforward computation based on our recursive
partitioning of the stream around where $F_2$ changes (see \cite{bciw15}
for details) shows that for any stream position $t$ and set $T_i$ we have created, 
$$d(t, T_i) = \min_{s \in T_i} \left ({\bf E}[|c(t) - c(s)|^2] \right )^{1/2} = O \left (\frac{F_2}{2^{2^i}} \right )^{1/2}.$$
Applying Theorem \ref{thm:chain}, we have
$${\bf E}[\sup_{t \in T} X_t ] \leq O(1) \sup_{t \in T} \sum_{i \geq 0} 2^{i/2} \left (\frac{F_2}{2^{2^i}} \right )^{1/2} = O(F_2^{1/2}).$$
This is exactly the same bound that the theory for random walks gave us earlier! (recall in that case $\sum_{i \neq i^*} f_i^2 < n$).

Using Gaussian processes has therefore allowed us to remove our earlier assumption that $f_i \in \{0,1\}$ for all 
$i \in [n] \setminus \{i^*\}$. The same random walk based algorithm will now work; however, we still need to assume
the $f_{i^*} = \Omega(\sqrt{F_2} \log n)$ in order to learn $\log n$ bits of information to identify $i^*$, as before. 
This is not satisfactory, as an $\ell_2$-heavy hitter only satisfies $f_i = \Omega(\sqrt{F_2})$ (recall we have assumed
$\phi$ and $\epsilon$ are constants), which is weaker than the $f_{i^*} = \Omega(\sqrt{F_2} \log n)$ that the above
analysis requires. 

\subsubsection{Amplification}
To remove the assumption that $f_{i^*} = \Omega(\sqrt{F_2} \log n)$, our work \cite{bciw15} designs what we call an 
``amplification'' procedure. This involves for $j = 1, 2, \ldots, O(\log \log n)$, independently 
choosing a pairwise independent hash function $h^j:[n] \rightarrow \{1, 2\}$. For each $j$, we as before maintain two counters
$c^j_1 = \sum_{i \mid h^j(i) = 1} g_j(i) \cdot f_i$ and $c^j_2 = \sum_{i \mid h^j(i) = 2} g_j(i) \cdot f_i$, where the $g_j(i)$ are independent
standard normal random variables. 

Applying the chaining inequality to each of the $O(\log \log n)$ counters created, we have that with large constant probability, in a constant fraction of the $O(\log \log n)$ pairs, both counters $c^j_1$ and $c^j_2$ will be bounded by $O(\sqrt{F_2})$ in magnitude. It follows that if $f_{i^*} \geq C \sqrt{F_2}$ for a sufficiently large constant $C > 0$ (which we can assume by first hashing the universe into $O(1)$ buckets before the streaming
algorithm begins), then in say, a $9/10$ fraction of pairs $j$, the counter $c^j_k$, $k \in \{1,2\}$, of larger magnitude will contain
$i^*$. Moreover, by Chernoff bounds, only a $\frac{1}{\log^c n}$ fraction of other $i \in [n]$ will hash to the larger counter in at least a $9/10$ fraction of such pairs, where $c > 0$ is a constant that can be made arbitrarily large by increasing the constant in the number $O(\log \log n)$ of pairs of counters created. Now the idea is to effectively run our previous algorithm only on items which hash to the heavier counter in at least a $9/10$ fraction of pairs. By definition, this will contain $i^*$, and now the expected second moment of the other items for which we run the algorithm on will be $F_2/\log^c n$, which effectively makes $f_{i^*} = \Omega(\sqrt{F_2} \log n)$, where $F_2$ is now measured with respect to the items for which we run the algorithm on. Now we can sequentially learn $O(\log n)$ bits of information about $i^*$ in our algorithm, as before. 

One thing to note about this approach is that after seeing a sufficiently large number of insertions of $i^*$, i.e., $\Theta(\sqrt{F_2})$ such insertions, then most of the pairs of counters will have the property that the larger counter (in absolute value) stays larger forever. This is due to the chaining inequality. This can be used to fix the itemset for which we run the algorithm on. In fact, this is precisely why this does not result in a $2$-pass algorithm, which one might expect since one does not know the itemset to run our algorithm on in advance. However, we always run the algorithm on whichever current itemset agrees with at least a $9/10$ fraction of the larger counters, and just accept the fact that in the beginning of the stream the bits we learn about $i^*$ are nonsense; however, after enough updates to $i^*$ have occurred in the stream then the counters ``fix'' themselves in the sense that the larger counter does not change. At this point the bits we learn about $i^*$ in our algorithm are the actual bits that we desire. At the end of the stream, we only look at a suffix of these bits to figure out $i^*$, thereby ignoring the nonsensical bits at the beginning of the stream. We refer the reader to \cite{bciw15} for more details.

\subsubsection{Derandomization}
The final piece of the algorithm is to account for the randomness used by the algorithm. We need to derandomize the counters, which use the theory of Gaussian processes to argue their correctness. We also cannot afford to maintain all of the hash functions that were used to learn specific bits of $i^*$ (which we need ad the end of the stream to figure out what $i^*$ is). 

To derandomize the Gaussian processes, we use a derandomized Johnson Lindenstrauss transform of Kane, Meka, and Nelson \cite{kmn11}. The rough idea is to first apply a Johnson-Lindenstrauss transform to the frequency vectors for which we take inner products with independent Gaussian random variables in our counters. This will reduce the dimension from $n$ to $O(\log n)$, for which we can then afford to take an inner product with fully independent Gaussian random variables. The nice thing about Johnson-Lindenstrauss transforms is that they preserve all the covariances up to a constant factor in our specific Gaussian process, and therefore we can use Slepian's Lemma (see \cite{bciw15} for details) to argue that the Gaussian process is roughly the same as before, since it is entirely determined by its covariances. Here the derandomized Johnson-Lindenstrauss transform of
\cite{kmn11} can be represented using only $O(\log n \log \log n)$ bits of space. Also,
instead of using Gaussian random variables, which require truncation, we can directly
use sign random variables ($+1$ with probability $1/2$, $-1$ with probability $1/2$), which
results in what are called Bernoulli processes, together with a comparison theorem
for Bernoulli processes and Gaussian processes. This enables us to avoid arguments
about truncating Gaussians. 

To derandomize the hash functions, we use Nisan's pseudorandom generator in a similar way that Indyk uses it for derandomizing his algorithms for norm estimation \cite{i06,n92}. Please see \cite{bciw15} for further details.

\subsection{Followup Work}

Very recently, in followup work by Braverman et al. \cite{bcinww16}, we improved the space bound further to 
the optimal $O(\log n)$ bits of space (for constant $\epsilon, \phi$). 
The high level idea of using 
Gaussian or Bernoulli processes is the same, but several additional insights
were needed. This involves both a new algorithm which we call {\sf BPTree},
which avoids the amplification step and Nisan's pseudorandom generator 
described above, 
as well as a better derandomization of
the Bernoulli processes using $O(1)$-wise independence. 
We refer the reader to that work for further details. 

\section{Conclusions}
We presented new algorithms for finding $\ell_1$-heavy hitters and $\ell_2$-heavy hitters in a data stream. We refer the reader to
the original papers cited above for further details. As these problems are inspired from applications in practice, it is very
interesting to see how the improved theoretical algorithms perform in practice. In ongoing work we are testing these algorithms in practice on real datasets. 

Another
interesting aspect is that the technique of using Gaussian processes in the $\ell_2$-heavy hitters algorithm has led to a number
of other improvements to data stream algorithms, including for example 
the ability to estimate the second moment $F_2$ at all times in a stream of insertions. Previously, given a stream of length $n$ and a universe of
size $n$, to estimate $F_2$ at all points in a stream up to a constant factor would require $\Theta(\log^2 n)$ bits of space, since it
takes $\Theta(\log n \log(1/\delta))$ bits to estimate it at a single point with failure probability $\delta$, 
and one needs to union bound over $n$ stream positions. Using Gaussian processes, \cite{bciw15} achieves only $O(\log n \log \log n)$ bits of space
for this task. It would be interesting to see if Gaussian processes are useful for other problems in data streams. 

\bibliographystyle{plain}
\bibliography{main}

\end{document}